\documentclass[a4paper,twoside,dvips,12pt]{article}
\usepackage[german,english]{babel}
\usepackage{./styles/mytimes}
\usepackage{epsfig}
\usepackage{picins}
\usepackage[centertags]{amsmath}
\usepackage{amssymb,amscd}
\usepackage{./styles/mcite}
\usepackage{./styles/mysidecaption}
\def\zero{{\scriptscriptstyle 0}}

\catcode`\@=11 
\@ifundefined{euro}%
{\providecommand{\Euro}{Euro\xspace}
 }%
{\providecommand{\Euro}{\euro\xspace}
 }
\catcode`\@=12 

\def\Z0{{Z^\zero}}
\def\eVdist{\kern-0.06667em}

\def\ev{{\,\text{e}\eVdist\text{V\/}}}

\def\gev{{\,\text{Ge}\eVdist\text{V\/}}}
\def\tev{{\,\text{Te}\eVdist\text{V\/}}}

\def\met{\,\text{m}}

\def\kHz{\,\text{kHz}}

\def\IP{{\rm I$\kern-0.01667em$P}\xspace}

\def\Ptlj{{\not{\kern-0.55ex P}}_t\ell j}
\def\Ptmiss{{\not{\kern-0.55ex P}}_t}

\def\rnge{{\,\text{--}\,}}

\newcommand{\nucl}[2]{\ensuremath{\stru{2.ex}{0.8ex}^{#2}{\rm #1}}}

\mathchardef\qsm=63
\mathchardef\pls=43
\mathchardef\mns=512
\mathchardef\plm=518
\mathchardef\eql=61
\mathchardef\smallleft=300
\mathchardef\smallright=301
\mathchardef\perslsh=47
\mathchardef\les=316
\mathchardef\gre=318
\mathchardef\leq=532
\mathchardef\grq=533
\chardef\usc=95
\chardef\til=126

\def\sqr#1#2#3{{\vcenter{\hrule height.#3ex\hbox{\vrule width.#2ex height#1ex
    \kern#1ex\vrule width.#3ex}\hrule height.#2ex}}}

\def\angleto{\vrule width.035em height2.1ex depth-.56ex\unskip\kern-.6ex\to}
\def\perchc#1{{\raise.4ex\hbox{$\mkern4mu#1{\it\perslsh}_
             {\mkern-5mu\scriptscriptstyle{{\rm o}\!{\rm o}}}^
             {\mkern-12.8mu\scriptscriptstyle{\rm o}}$}}}

\catcode`\@=11 
\def\parenbar{\mathpalette\p@renb@r}
\def\p@renb@r#1#2{\vbox{%
  \ifx#1\scriptscriptstyle \dimen@.7em\dimen@ii.2em\else
  \ifx#1\scriptstyle \dimen@.8em\dimen@ii.25em\else
  \dimen@1em\dimen@ii.4em\fi\fi \offinterlineskip
  \ialign{\hfill##\hfill\cr
    \vbox{\hrule width\dimen@ii}\cr
    \noalign{\vskip-.3ex}%
    \hbox to\dimen@{$\mathchar300\hfil\mathchar301$}\cr
    \noalign{\vskip-.3ex}%
    $#1#2$\cr}}}
\catcode`\@=12 

\newbox\struttbox
\setbox\struttbox=\hbox{\vrule height1.65ex depth.485ex width0pt}
\def\strutt{\relax\ifmmode\copy\struttbox\else\unhcopy\struttbox\fi}
\def\stru#1#2{\relax\ifmmode\hbox{\vrule height#1 depth#2 width0pt}
\else\vrule height#1 depth#2 width0pt\fi}
\def\uline#1{$\underline{\hbox{#1\strutt}}$}
\def\ronum#1{\uppercase\expandafter{\romannumeral#1}}
\def\ronuml#1{\expandafter{\romannumeral#1}}

\def\cbk{\kern-0.5em}
\newcommand{\pcite}[1]{{\protect\cite{#1}}}

\newcommand{\linebox}[2][3.ex]{\uline{\hbox to #2{\stru{#1}{0.pt}\hfil}}}

\newcounter{seqnum}
\DeclareMathAlphabet{\mathbf}{OT1}{cmr}{bx}{n}

\DeclareMathAlphabet{\mathbfs}{OT1}{lcmss}{bx}{sl}

\newlength\listtextwidth

\catcode`\@=11 
\newlength{\@tabfninsert}
\newlength{\@tabfnwidth}
\newcommand{\tabfootnote}[2]{%
  \setlength{\@tabfninsert}{0.8em}
  \setlength{\@tabfnwidth}{\textwidth}
  \addtolength{\@tabfnwidth}{-\@tabfninsert}
  \addtolength{\@tabfnwidth}{-0.4em}
  \noindent\makebox[\@tabfninsert][r]{\footnotesize$^{#1}$\hfil}\hfill%
  \parbox[t]{\@tabfnwidth}{\footnotesize #2\hfill}}
\catcode`\@=12 
\newcommand{\boldarrayrulewidth}{1pt}

\catcode`\@=11 
\let\tab@penalty\relax
\newcount\tab@state
\def\bcline#1{%
  \noalign{\kern-.5\arrayrulewidth\tab@penalty}%
  \omit%
  \global\tab@state\@ne%
  \ranges\bcline@i{#1}%
  \cr%
  \noalign{\kern-.5\arrayrulewidth\tab@penalty}%
}
\def\bcline@i#1#2{%
  \ifnum#1<\tab@state\relax%
    \tab@@cr%
    \noalign{\kern-\arrayrulewidth\tab@penalty}%
    \omit%
    \global\tab@state\@ne%
  \fi%
  \@whilenum\tab@state<#1\do{%
    \hfil\tab@@tab@omit%
    \global\advance\tab@state\@ne%
  }%
  \ifnum\tab@state>\@ne%
    \kern-\arrayrulewidth%
  \fi%
  \@whilenum\tab@state<#2\do{%
    \tab@@span@omit%
    \global\advance\tab@state\@ne%
  }%
  \leaders\hrule\@height\boldarrayrulewidth\hfill%
}
\def\ranges#1#2{%
  \gdef\ranges@temp{#1}%
  \begingroup%
  \ranges@i#2 \q@delim%
}
\def\ranges@i{%
  \@ifnextchar\q@delim\ranges@done{\afterassignment\ranges@ii\count@}%
}
\def\ranges@ii{%
  \@ifnextchar-\ranges@iii{\ranges@do\count@\count@\ranges@v}%
}
\def\ranges@iii-{\afterassignment\ranges@iv\@tempcnta}
\def\ranges@iv{\ranges@do\count@\@tempcnta\ranges@v}
\def\ranges@v{%
  \@ifnextchar,%
    \ranges@vi%
    {%
      \@ifnextchar\q@delim%
        \ranges@done%
        {\tab@err@range\ranges@vi,}%
    }%
}
\def\ranges@vi,{\afterassignment\ranges@ii\count@}
\def\ranges@do#1#2{%
  \ifnum#1>#2\else%
    \expandafter\endgroup%
    \expandafter\ranges@temp%
    \expandafter{%
    \the\expandafter#1%
    \expandafter}%
    \expandafter{%
    \the#2%
    }%
    \begingroup%
  \fi%
}
\def\ranges@done\q@delim{\endgroup}
\def\ifinrange#1#2{%
  \@tempswafalse%
  \count@#1%
  \ranges\ifinrange@i{#2}%
  \if@tempswa%
    \expandafter\@firstoftwo%
  \else%
    \expandafter\@secondoftwo%
  \fi%
}
\def\ifinrange@i#1#2{%
  \ifnum\count@<#1 \else\ifnum\count@>#2 \else\@tempswatrue\fi\fi%
}
\def\tab@@cr{\cr}
\def\tab@@tab@omit{&\omit}
\def\tab@@span@omit{\span\omit}
\def\tab@checkrule#1{%
  \count@#1\relax%
  \expandafter\ifinrange%
  \expandafter\count@%
  \expandafter{\tab@xcols}%
    {\tab@checkrule@i}%
    {}%
}
\def\bhline{\noalign{\ifnum0=`}\fi\hrule \@height  
\boldarrayrulewidth \futurelet \@tempa\@xhline}
\def\@xhline{\ifx\@tempa\hline\vskip \doublerulesep\fi
      \ifnum0=`{\fi}}
\catcode`\@=12 

\catcode`\@=11 
\newcounter{pict@width}
\newcounter{pict@height}
\newlength{\pict@scale}
\newcommand{\psfigadd}[4]{%
\setcounter{pict@width}{1*\ratio{#2+\pict@scale/2}{\pict@scale}}
\setcounter{pict@height}{1*\ratio{#3+\pict@scale/2}{\pict@scale}}
\setlength{\unitlength}{\pict@scale}
\hbox to #2{\hspace{-\fill}\begin{picture}(\thepict@width,\thepict@height)
\put(0,0){\psfig{figure=#1,width=#2,height=#3,clip=}}
\SetScale{0.283466457}
\SetWidth{1.763889}
{#4}
\end{picture}}
}
\newcounter{pict@widthfst}
\newcounter{pict@widthscd}
\newcounter{pict@widthtot}
\newcommand{\psfigaddtwo}[7]{%
\setcounter{pict@widthfst}{1*\ratio{#2+\pict@scale/2}{\pict@scale}}
\setcounter{pict@widthscd}{1*\ratio{#2+#4+\pict@scale/2}{\pict@scale}}
\setcounter{pict@widthtot}{1*\ratio{#2+#4+#6+\pict@scale/2}{\pict@scale}}
\setcounter{pict@height}{1*\ratio{#3+\pict@scale/2}{\pict@scale}}
\setlength{\unitlength}{\pict@scale}
\hbox{\hspace{-\fill}\begin{picture}(\thepict@widthtot,\thepict@height)
\put(0,0){\psfig{figure=#1,width=#2,height=#3,clip=}}
\put(\thepict@widthscd,0){\psfig{figure=#5,width=#6,height=#3,clip=}}
\SetScale{0.283466457}
\SetWidth{1.763889}
{#7}
\end{picture}}
}
\newcommand{\psfigror}[4]{%
\setcounter{pict@width}{1*\ratio{#2+\pict@scale/2}{\pict@scale}}
\setcounter{pict@height}{1*\ratio{#3+\pict@scale/2}{\pict@scale}}
\setlength{\unitlength}{\pict@scale}
\hbox{\begin{picture}(\thepict@width,\thepict@height)
\put(0,\thepict@height){\psfig{figure=#1,width=#3,height=#2,clip=,angle=270}}
\SetScale{0.283466457}
\SetWidth{1.763889}
{#4}
\end{picture}}
}
\newcommand{\psfigrol}[4]{%
\setcounter{pict@width}{1*\ratio{#2+\pict@scale/2}{\pict@scale}}
\setcounter{pict@height}{1*\ratio{#3+\pict@scale/2}{\pict@scale}}
\setlength{\unitlength}{\pict@scale}
\hbox{\begin{picture}(\thepict@width,\thepict@height)
\put(0,0){\psfig{figure=#1,width=#3,height=#2,clip=,angle=90}}
\SetScale{0.283466457}
\SetWidth{1.763889}
{#4}
\end{picture}}
}
\catcode`\@=12 

\begin{document}
\title{ 
\vspace*{-1.cm}
\rightline{\normalsize\bf FAU-PI1-05-01}
\vspace*{1.8cm}
Neutrino Telescopy in the Mediterranean Sea}
\author{Ulrich F.~Katz\\
University of Erlangen, Physikalisches Institut, Germany
}
\maketitle
\begin{abstract} 
The observation of high-energy extraterrestrial neutrinos is one of the most
promising future options to increase our knowledge on non-thermal processes in
the universe. Neutrinos are e.g.\ unavoidably produced in environments where
high-energy hadrons collide; in particular this almost certainly must be true in
the astrophysical accelerators of cosmic rays, which thus could be identified
unambiguously by sky observations in ``neutrino light''. On the one hand,
neutrinos are ideal messengers for astrophysical observations since they are not
deflected by electromagnetic fields and interact so weakly that they are able to
escape even from very dense production regions and traverse large distances in
the universe without attenuation. On the other hand, their weak interaction
poses a significant problem for detecting neutrinos. Huge target masses up to
gigatons must be employed, requiring to instrument natural abundances of media
such as sea water or antarctic ice. The first generation of such neutrino
telescopes is taking data or will do so in the near future, while the
second-generation projects with cubic-kilometre size is under construction or
being prepared. This report focuses on status and prospects of current (ANTARES,
NEMO, NESTOR) and future (KM3NeT) neutrino telescope projects in the
Mediterranean Sea.
\end{abstract}

\section{Current neutrino telescope projects in the Mediterranean Sea}
\label{sec-cur}

World-wide, two neutrino telescopes (AMANDA at the South Pole
\cite{app:13:1,prl:92:071102-1} and one in Lake Baikal
\cite{app:7:263,*pan:63:951}) are taking data, two are under construction in the
Mediterranean Sea (ANTARES \cite{astro-ph-9907432,misc:ant:tdr}, NESTOR
\cite{ncim:24c:771,proc:icrc:2003:1377}), and the cubic-kilometre telescope
IceCube \cite{misc:icube:pdd,*misc:icrc01:ic} is being installed at the South
Pole. Preparatory work for a corresponding installation in the Mediterranean Sea
is being performed in the R\&D project NEMO \cite{npps:143:359,*misc:nemo-home};
from early 2006 on, all groups involved in the current Mediterranean projects
will join into a 3-year EU-funded Design Study towards the future km$^3$-scale
neutrino telescope in the Northern hemisphere (KM3NeT)
\cite{misc:km3net:homepage}.

\subsection{Detection principle}
\label{sec-cur-det}

Interactions of neutrinos with target material in the neutrino detector or its
vicinity produce charged secondary particles with velocities exceeding the speed
of light in water or ice, which therefore radiate \v Cerenkov light. This light
is detected by an array of photomultipliers placed deep below the surface. The
range of neutrino energies for which neutrino telescopes are sensitive is
limited by this detection method to some $10\gev$ at its lower end, while at
energies beyond roughly $10^{17}\ev$ the neutrino flux is expected to fade below
detection thresholds even for future giant detectors.

From the photomultiplier positions, the arrival time of the light (measured to
nanosecond precision) and the signal amplitudes, the direction and energy of the
incoming neutrino are reconstructed. The achievable resolutions depend on the
reaction type: Charged-current reactions of muon neutrinos\footnote{In the
following, the term {\it neutrino} is generically used to denote both neutrinos
and antineutrinos}, $\nu_\mu N\to\mu X$, produce high-energy muons with a range
of up to several kilometres in water or ice; the detection of these muons allows
for a precise reconstruction of the neutrino direction\footnote{When referring
to angular resolution in the following, this event type is assumed.} (resolution
in water better than $0.3^\circ$ for neutrino energies $E_\nu\gtrsim10\tev$) and
an estimate of the neutrino energy accurate to within a factor of 2 for
$E_\nu\gtrsim1\tev$. Due to the good angular resolution and the increased
sensitivity resulting from the large muon range, neutrino telescopes are
predominantly optimised for this reaction type. On the other hand,
charged-current reactions of electron or tau neutrinos, $\nu_{e,\tau} N\to
(e,\tau)X$, and neutral-current reactions, $\nu_x N\to\nu_xX$, produce hadronic
and/or electromagnetic particle cascades ({\it showers}) which act as localised
sources of intense \v Cerenkov light. Such reactions occurring inside the
instrumented volume allow for a rather precise measurement of the shower energy,
with an angular resolution degraded to several degrees in water, and even worse
in ice.

\begin{figure}[hbt]
  \sidecaption
  \epsfig{file=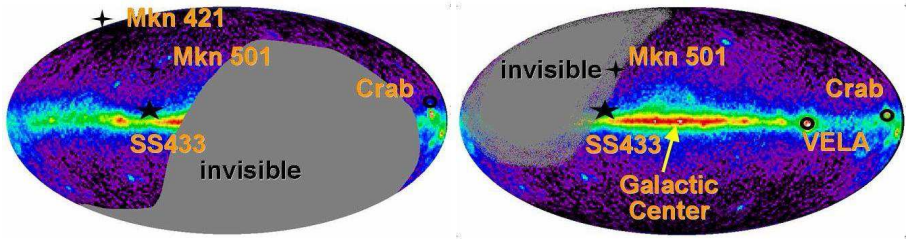,width=12.cm}
  \caption{\protect\raggedright
           Field of view of a neutrino telescope at the South Pole (left) and in
           the Mediterranean (right), given in galactic coordinates. A
           $2\pi$-downward sensitivity is assumed; the gray regions are then
           invisible. Indicated are the positions of some candidate neutrino
           sources.}
  \label{fig-fov}
\end{figure}

In order to shield the experiments against background daylight and muons
originating from cosmic ray interactions in the upward-hemisphere atmosphere
({\it atmospheric muons}), they are located in a depth of several kilometres. 
Yet, for most of the abovementioned energy range the atmospheric muon background
is prohibitive for observing neutrinos arriving from above. Therefore, the field
of view of neutrino telescopes is the downward hemisphere; observing the
Southern sky including the Galactic Centre hence requires an experiment in the
Earth's Northern hemisphere. A comparison of the fields of view from the South
Pole and the Mediterranean Sea is shown in Fig.~\ref{fig-fov}. At highest
energies beyond roughly $10^{15}\ev$ the atmospheric muon flux fades away and
the view opens to the upper hemisphere; at the same time, the downward view
becomes obscured by the fact that, due to the increase of the neutrino cross
section with energy, the Earth becomes opaque even for neutrinos.

\subsection{General conditions in sea water}
\label{sec-cur-gen}

The major challenges in constructing deep-sea neutrino telescopes are the high
pressure of several $100\,$bar; the uncontrollable environment with currents,
sedimentation and background light from $\nucl{K}{40}$ decays and bioluminescent
organisms; the chemically aggressive environment reducing the selection of
suited materials basically to titanium, glass and certain plastics. A further
aspect of this difficult environment is that the deployment and maintenance
operations, employing surface vessels and manned or remotely-operated deep-see
submersibles, are expensive and weather-dependent, thus maximising the need for
high operation stability.

In addition to requirements implied by pressure and material choices, the
parameters that affect the detector design most strongly are the water
transparency and the background light. In the Mediterranean deep-sea
environment, the absorption length of blue light is close to $60\met$, implying
distances of detection units of this order or less. The light scattering length
exceeds $200\met$, resulting in a very good angular resolution, by far superior
to the one achievable in polar ice where scattering is much stronger. The
presence of $\nucl{K}{40}$ causes a steady background of single-photon signals,
amounting to about $30\rnge40\kHz$ per 10-inch photomultiplier; bioluminescence
light causes additional steady and burst-like background components. This
situation requires a high data transmission bandwidth as well as stringent
coincidence triggers exploiting the good timing resolution.

\subsection{ANTARES}
\label{sec-cur-ant}

\begin{figure}[hbt]
  \sidecaption
  \epsfig{file=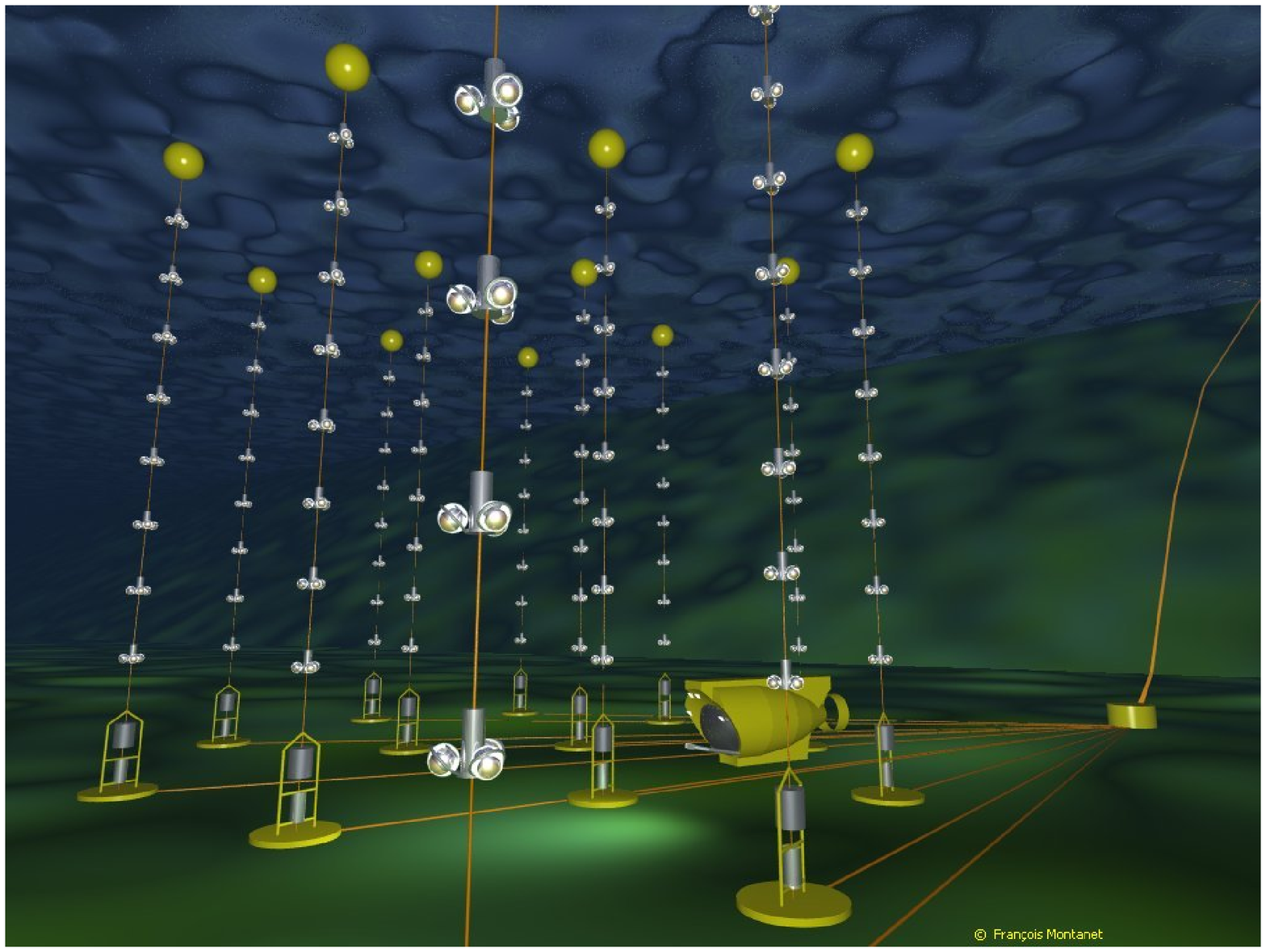,height=6.cm}
  \epsfig{file=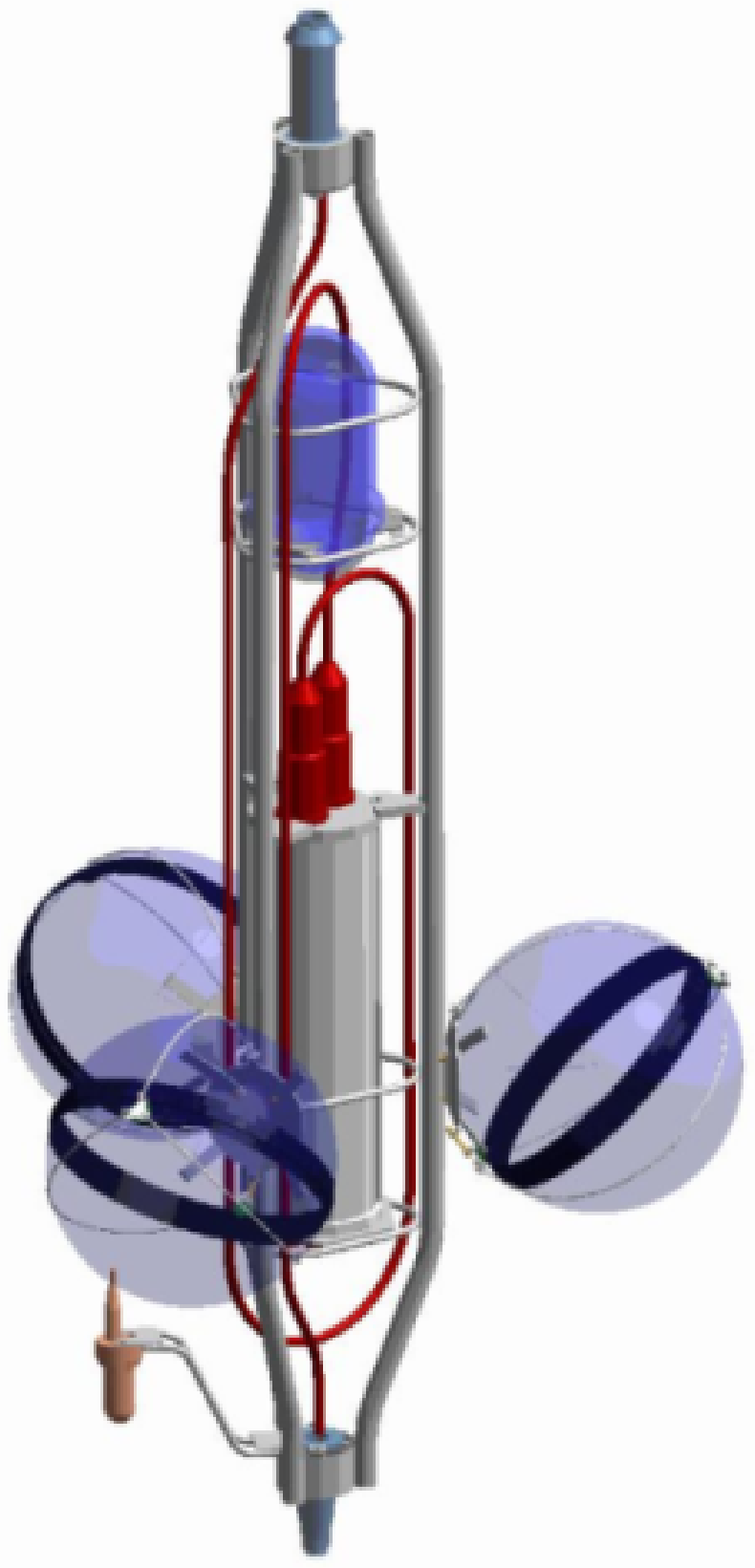,height=6.cm}\kern-1.mm
  \caption{\protect\raggedright
           An artist's view of the ANTARES detector (left, not to scale) and a
           schematic view of a storey (right) with the three glass spheres for
           the photomultipliers, an optical beacon for time calibration (blue)
           and a hydrophone for position measurement (below front sphere).}
  \label{fig-ant}
\end{figure}

The ANTARES neutrino telescope \cite{astro-ph-9907432} is currently under
construction off the French Mediterranean coast near Toulon. It will be situated
in $2500\met$ depth and will consist of 12~lines ({\it ``strings''}) that are
anchored to the sea bed at distances of about $70\met$ from each other and kept
vertical by buoys (see Fig.~\ref{fig-ant}). Each string is equipped with 75
optical modules (OMs) \cite{nim:a484:369} arranged in triplets ({\it storeys})
sustained by titanium frames that also support water-tight titanium containers
for the electronic components. The OMs are glass spheres housing one 10-inch
photomultiplier each, directed at an angle of $45^\circ$ towards the sea bed. 
The storeys are spaced at a vertical distance of $14.5\met$ and are
interconnected with an electro-optical-mechanical cable supplying the electrical
power and the control signals and transferring the data to the string bottom. 
Submersible-deployed electro-optical link cables connect the strings to the {\it
junction box (JB)}, which acts as a fan-out between the main electro-optical
cable to shore and the strings. Each string carries optical beacons for timing
calibration and acoustic transponders used for position measurements. The
detector will be complemented by an {\it instrumentation line} supporting
devices for measurements of environmental parameters as well as tools used by
other scientific communities, such as e.g.\ a seismometer.

The main electro-optical cable and the junction box are installed and
operational since 2002. Two prototype strings, one with 5 optical storeys and
one with auxiliary instrumentation, were deployed, connected and operated in
2003. Based on the results of these prototypes, the design was finalised and
scrutinised using two further test strings deployed in 2005. The full
functionality of the detector has been verified to design specifications. 
Currently, the first full detector line is awaiting deployment; the detector
installation is expected to be completed by 2007.

For a detailed summary of the results of the ANTARES test deployments see
\cite{misc:antares-status:1,*misc:antares-status:2}.

\subsection{NESTOR}
\label{sec-cur-nes}

\begin{figure}[hbt]
\begin{minipage}[b]{0.55\textwidth}
  \epsfig{file=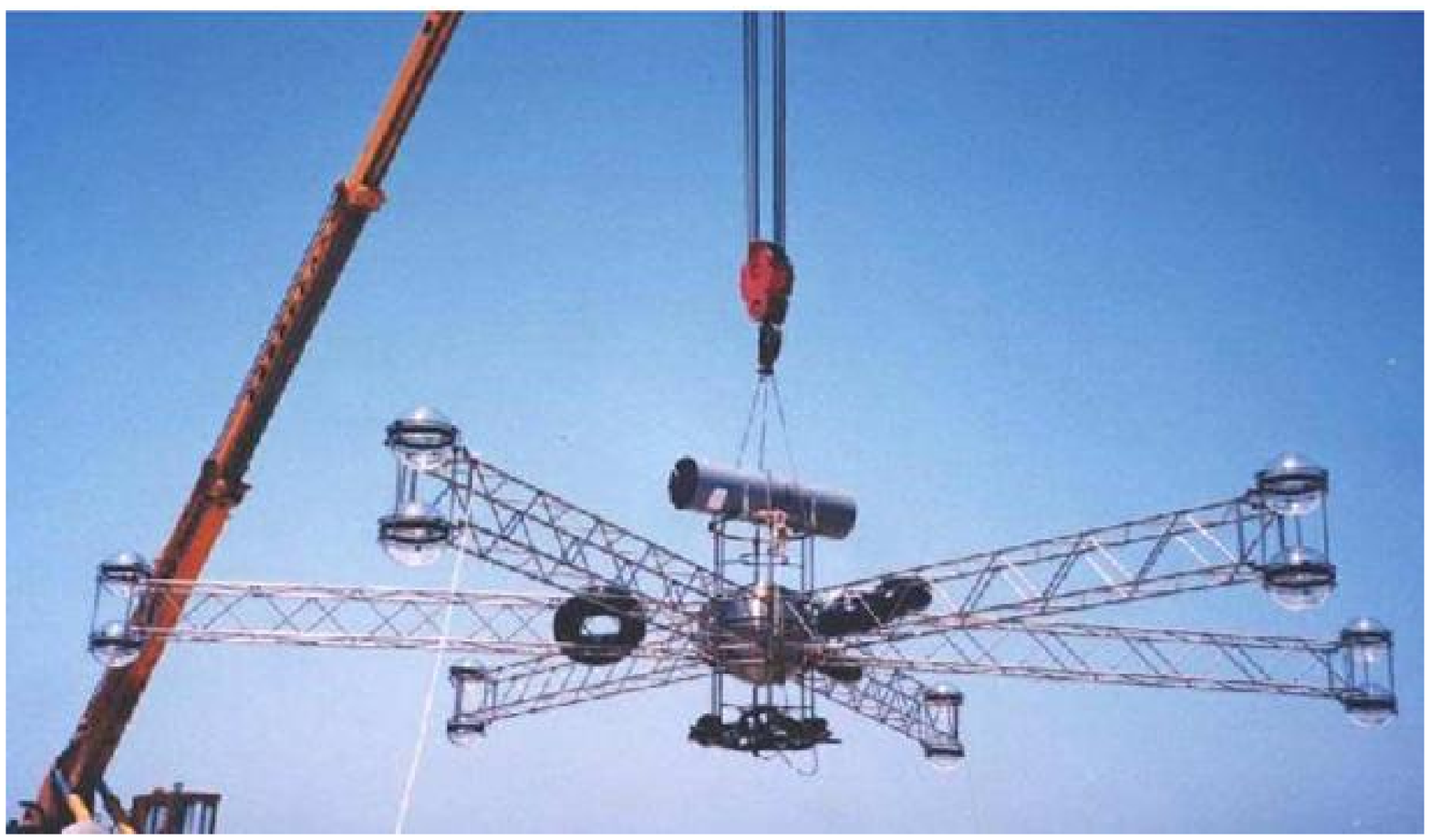,height=5.5cm}
  \caption{\protect\raggedright
           Reduced-size NESTOR floor during preparation for deployment.} 
  \label{fig-nes-flo}
\end{minipage}\hfil
\begin{minipage}[b]{0.42\textwidth}
  \epsfig{file=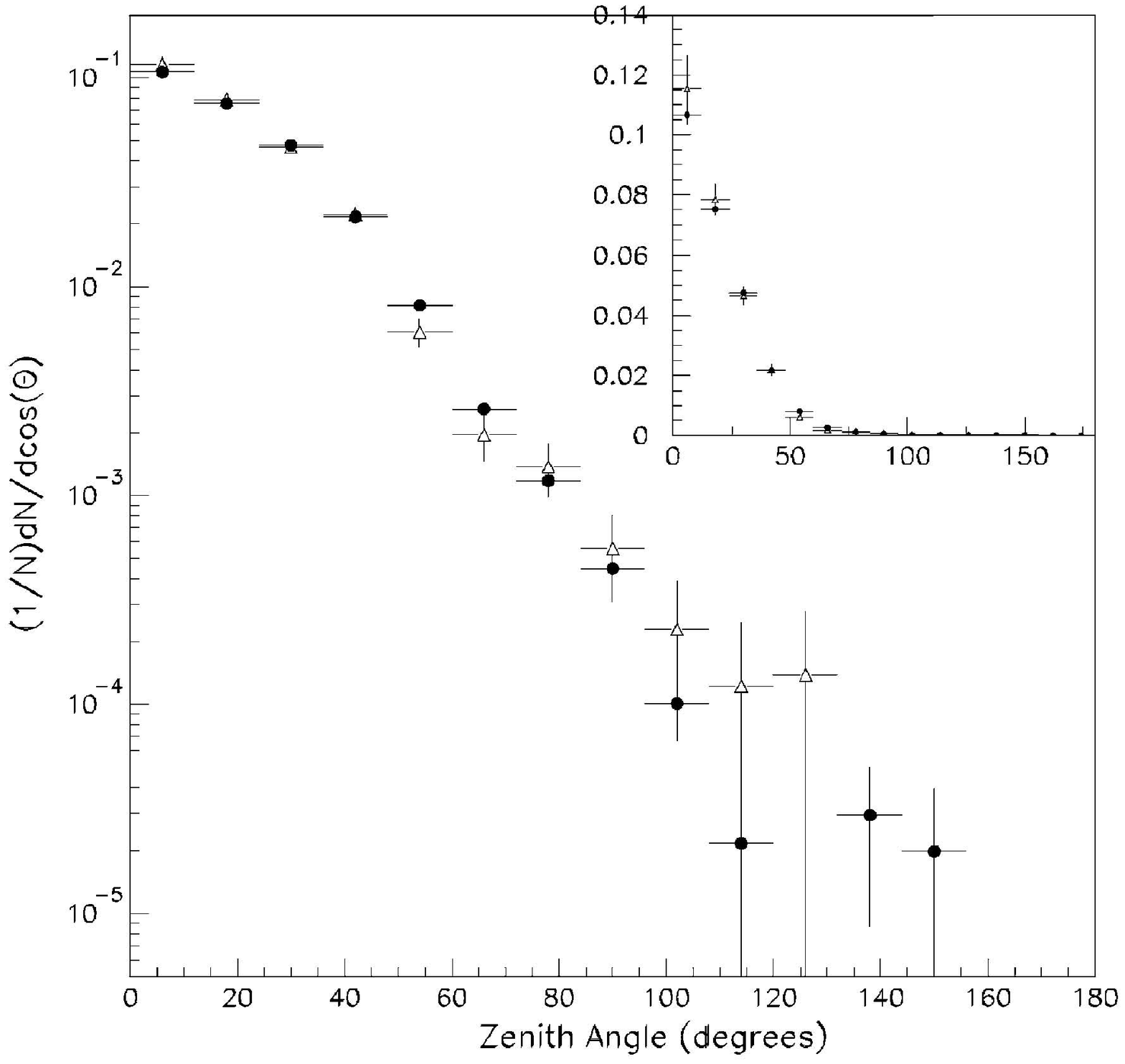,height=6.6cm}
  \caption{\protect\raggedright
           Zenith angle distribution of atmospheric muons measured by NESTOR
           during the test deployment (triangles), compared to the result of a
           simulation (filled circles). The insert shows the same data with a
           linear vertical scale. The figure has been taken from
           \pcite{nim:a552:420}.}
  \label{fig-nes-ang}
\end{minipage}
\end{figure}

The site selected for the NESTOR neutrino telescope is off Pylos at the West
coast of the Peloponnese, at a depth of $3800\met$. The NESTOR design is based
on rigid, hexagonal star-like structures ({\it floors}) with a diameter of
$32\met$, carrying 6 pairs of upward- and downward-looking photomultipliers each
as well as a titanium sphere for the readout electronics in the centre (see
Fig.~\ref{fig-nes-flo}). 12~floors will be connected vertically at a distance of
$30\met$ to form a {\it tower}. The deployment operations are performed by
lifting the existing structure to the surface, connecting the new module(s) and
redeploying the extended set-up, thus avoiding the use of submersibles.

In 2003, a single floor of reduced size has been deployed, connected to the
cable to shore and operated for more than a month \cite{nim:a552:420}. In this
time, more than 2~million 4- or higher-fold coincidence triggers have been
collected \cite{app:23:377}. These data allowed the NESTOR collaboration to
reconstruct the angular distribution of the atmospheric muons and to compare the
result to simulations and previous measurements. The good agreement found (cf.\
Fig.~\ref{fig-nes-ang}) confirms that the functionality of the detector complies
with the specifications and that a detailed level of detector understanding has
been reached.

\subsection{NEMO}
\label{sec-cur-nem}

\piccaption{\protect\raggedright Schematic view of a flexible NEMO tower.
            \label{fig-nem}}
\parpic(4.cm,6.5cm)(0pt,6.5cm)[r][t]{%
\epsfig{file=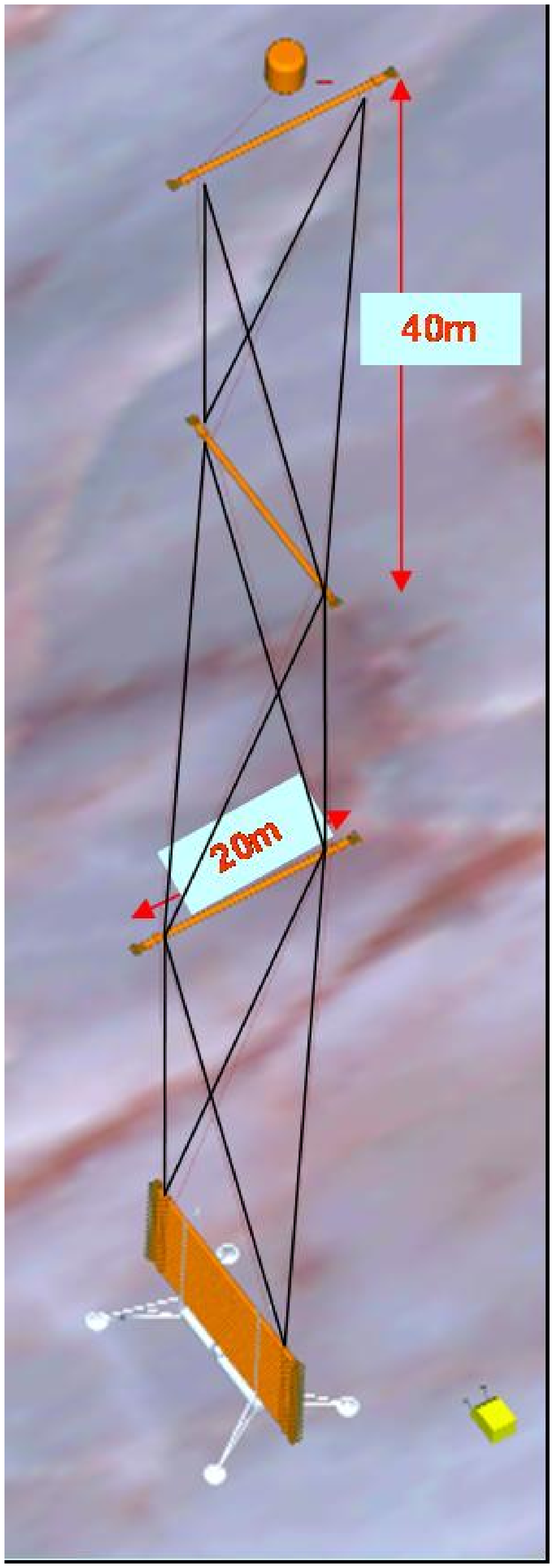,width=3.8cm,height=8.cm}
}
In the framework of the Italian R\&D project NEMO, a candidate site for a future
km$^3$-scale detector has been identified at a depth of $3340\met$ off the East
coast of Sicily near Capo Passero, and new solutions for various detector
components have been developed. Amongst these is a new design of a mechanical
structure, consisting of $20\met$-long rigid arms connected to each other by
ropes and kept vertical by a buoy. The ropes form a tetrahedral structure,
sustaining successive arms orthogonally to each other at a distance of $40\met$
(see Fig.~\ref{fig-nem}). Each arm carries 2 pairs of upward- and
downward-looking photomultipliers. One advantage of this flexible tower
structure is that a tower can be deployed folded into a compact structure which
unfurls when released after reaching the sea bottom. A further NEMO development
is a composite junction box, consisting of an inner, pressure-resistant steel
vessel embedded in an oil-filled plastic tank, thus separating the resistance
against pressure and salt water.

For assessing the newly developed components, a test site at a depth of
$2000\met$ has been identified and connected to the shore station by an
electro-optical cable. Within the forthcoming year, it is foreseen to deploy and
connect to this cable a junction box and a prototype of a flexible tower.
\section{Towards a km$^\mathbf{3}$ neutrino telescope in the Mediterranean Sea}
\label{sec-aim}

\looseness=-1
Already in 2002 the {\it High Energy Neutrino Astronomy Panel (HENAP)} of the
PaNAGIC\footnote{Particle and Nuclear Astrophysics and Gravitation International
Committee} Committee of \hbox{IUPAP}\footnote{International Union of Pure and
Applied Physics} has concluded that ``a km$^3$-scale detector in the Northern
hemisphere should be built to complement the IceCube detector being constructed
at the South Pole'' \cite{misc:henap:2002}; one major argument in favour of this
effort is the coverage of the Southern sky including the central part of the
Galactic plane (cf.\ Fig.~\ref{fig-fov} and Sect.~\ref{sec-phy}). This has
triggered a joint activity of the groups involved in the Mediterranean neutrino
telescopes towards establishing a common future project. The EU-funded KM3NeT
Design Study (see below) has been approved to prepare this project. 
Concurrently, the {\it European Strategy Forum for Research Infrastructures
(ESFRI)} has included the KM3NeT neutrino telescope in its {\it List of
Opportunities}
\cite{misc:esfri-loo:2005}, thus assigning high priority to this project.

\subsection{The KM3NeT Design Study}
\label{sec-aim-km3}

Even though making use of the experience and expertise gained in the current
projects, a major R\&D program has to be executed to arrive at a cost-effective
design for a km$^3$-scale deep-sea neutrino telescope, optimised for scientific
sensitivity, fast and secure production and installation, stable operation and
maintainability. The KM3NeT Design Study \cite{misc:km3net:homepage} will
address these issues in a 3-year program, with a 20\,M\Euro budget, of which
9\,M\Euro are provided by the EU. Participants are 29 particle/astroparticle and
7 sea science/technology institutes from altogether 8 European countries,
coordinated by the University of Erlangen.

Amongst the major issues to be studied and decided upon are the mechanical
structures, the choice of the photo-sensors, the readout, data acquisition and
online filter methods, the deep-sea infrastructure and deployment techniques;
for all of these, detailed simulation work will be necessary.

The main deliverable of the Design Study is a {\it Technical Design Report
(TDR)}, laying the foundation for funding negotiations and concrete project
preparation. The vision of the proponents is that KM3NeT will be a pan-European
research infrastructure, giving open access to the neutrino telescope data,
allowing to assign ``observation time'' to external users by adapting the online
filter algorithms to be particularly sensitive in predefined directions, and
also providing access to long-term deep-sea measurements for the marine science
communities.

\subsection{Timelines towards realisation}
\label{sec-aim-tim}

The KM3NeT Design Study will last until January 2009. Thereafter, a phase of
funding negotiations and construction preparation has to be foreseen, lasting
1--2 years. This phase might be supported within the European FP7 program. If
the decision to realise the KM3NeT infrastructure is taken in this phase,
installation could start as early as 2010 and be concluded in 2012. First data
would thus become available in 2011, concurrently with data from the IceCube
telescope which will be ready by then.

\section{Physics with neutrino telescopes}
\label{sec-phy}

After the above summary of the status and developments of the Mediterranean
neutrino telescopes, we will now highlight some of the physics issues related to
the interpretation of their data.

The lower part of the energy range defined in Sect.~\ref{sec-cur-det} is
dominated by the flux of {\it atmospheric neutrinos} (cf.\ Fig.~\ref{fig-difl}),
produced in reactions of cosmic rays with the Earth's atmosphere. The
atmospheric neutrinos establish a highly useful calibration source for the
detectors, but at the same time form an irreducible background in searching for
extraterrestrial neutrinos.

There are three basic search strategies to fight this background:
\begin{enumerate}
\item
Neutrinos from specific astrophysical objects (called {\it point sources})
produce excess signals associated to particular celestial coordinates and can
thus be identified on a statistical basis.
\item
Cosmic neutrinos are in general expected to have a much harder energy spectrum
than the atmospheric neutrinos. Neutrinos not associated to specific point
sources ({\it diffuse flux}) can thus be identified, again on a statistical
basis, by analysing the energy distribution of registered neutrino events.
\item
Exploiting coincidences in time and/or direction of neutrino events with
observations by telescopes (e.g.\ in the radio, visible, X-ray or gamma regimes)
and possibly also by cosmic ray detectors can be used to optimise search
strategies and to increase drastically the significance of observations of
transient sources ({\it multimessenger method}).
\end{enumerate}

The various astro- and particle physics questions to be addressed with the
resulting data have been summarised e.g.\ in \cite{astro-ph-0503122} and
references therein. Here, we will focus on a few central topics and a recent
development:
\begin{enumerate}
\item
\uline{Neutrinos from galactic shell-type supernova remnants:}\\
The shock waves developing when supernovae ejecta hit the interstellar medium
are prime candidates for hadron acceleration through the Fermi mechanism. Recent
observations of gamma rays up to energies of about $40\tev$ from two shell-type
supernova remnants in the Galactic plane (RX\,J1713.7-3946 and RX\,J0852.0-4622)
\cite{astro-ph-0511678,aa:437:l7} with the H.E.S.S.\ \v Cerenkov telescope
support this hypothesis and disfavour explanations of the gamma flux by purely
electromagnetic processes. The detection of neutrinos from these sources would,
for the first time, identify unambiguously specific cosmic accelerators. Note
that this is only possible with Northern-hemisphere neutrino telescopes which,
in contrast to the South Pole detectors, cover the relevant part of the Galactic
plane in their field of view (cf.\ Sect.~\ref{sec-cur-det}).

The expected event rates can be estimated using the rough assumption that the
muon neutrino and gamma fluxes are in relation $\phi_{\nu_\mu}/\phi_\gamma=1/2$,
taking into account the relative production probabilities of charged and neutral
pions, their decay chains and neutrino oscillations. Preliminary calculations
indicate that the first-generation Mediterranean neutrino telescopes may have a
chance to observe a few events, whereas a significantly larger signal is
expected in a future cubic-kilometre set-up; a tentative estimate of the neutrino
sky map of RX\,J0852.0-4622 after 5~years of data taking with KM3NeT is shown in
Fig.~\ref{fig-km3hess}.

\begin{figure}[hbt]
  \sidecaption
  \kern1.cm\epsfig{file=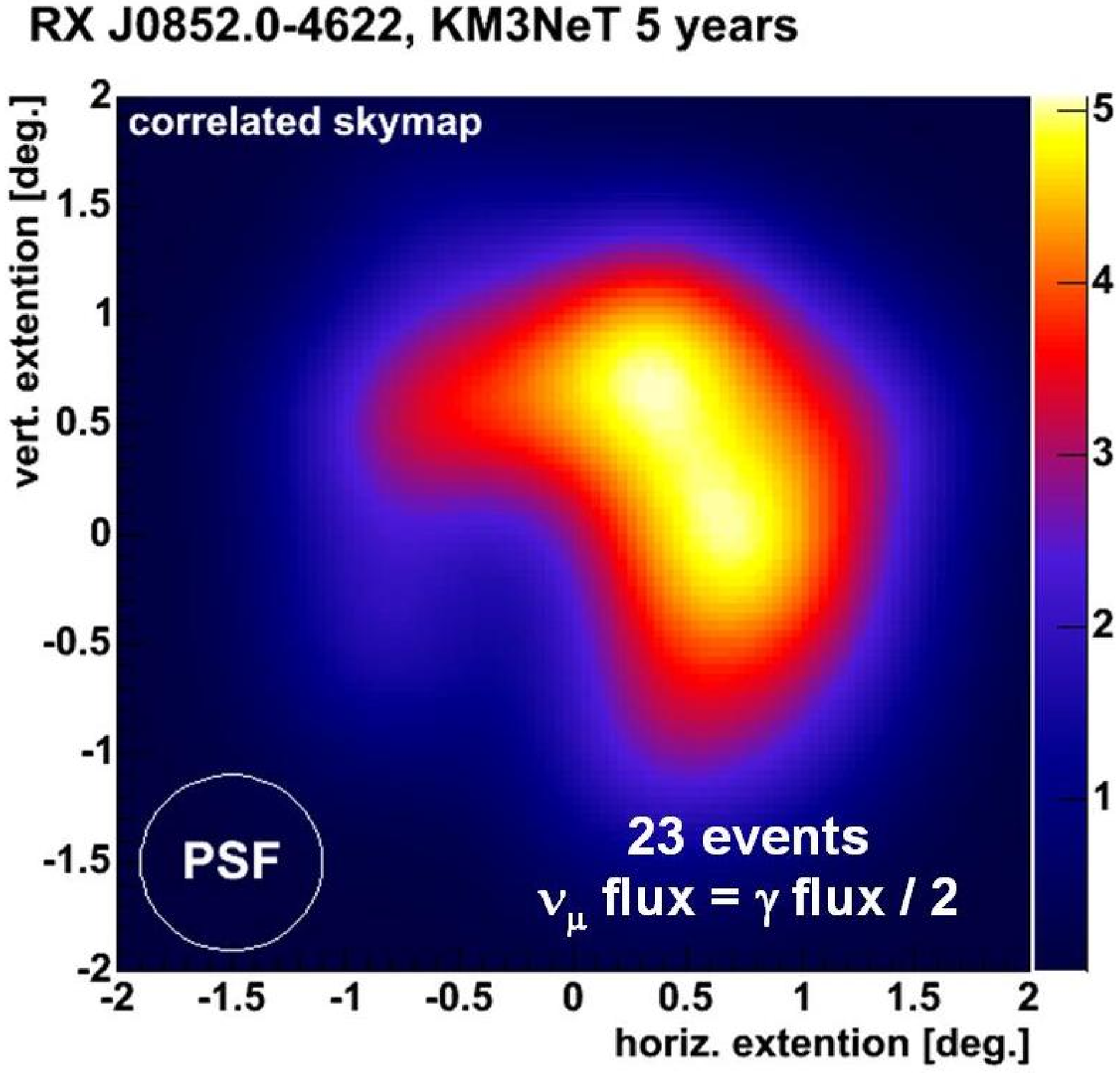,height=7.5cm}
  \caption{\protect\raggedright
           A skymap of the simulated neutrino signal from RX\,J0852.0-4622 as seen
           by a km$^3$-scale neutrino telescope in the Mediterranean Sea after 5
           years of data taking. The assumption $\phi_{\nu_\mu}/\phi_\gamma=1/2$
           has been used for the simulation. The background of atmospheric
           neutrinos, not included in the plot, amounts to a few events and can
           be efficiently eliminated by adjusting the lower energy cut without
           affecting significantly the signal. The circle in the lower left corner
           indicates the average angular resolution (point spread function).}
  \label{fig-km3hess}
\end{figure}

\item
\uline{The diffuse neutrino flux}\\
The sensitivity of current and future experiments is compared to various
predictions of diffuse neutrino fluxes in Fig.~\ref{fig-difl} (following
\cite{jp:g29:843,arevns:g29:843}). Whereas some of the models are already now
severely constrained by the data, others require km$^3$-size neutrino telescopes
for experimental assessment and potential discoveries. The measurement of the
diffuse neutrino flux would allow for important clues on the properties of the
sources, on their cosmic distribution, and on more exotic scenarios such as
neutrinos from decays of topological defects or superheavy particles ({\it
top-down scenarios}).

\begin{figure}[htb]
  \strut\kern1.cm
  \begin{minipage}{17.5cm}
  \epsfig{file=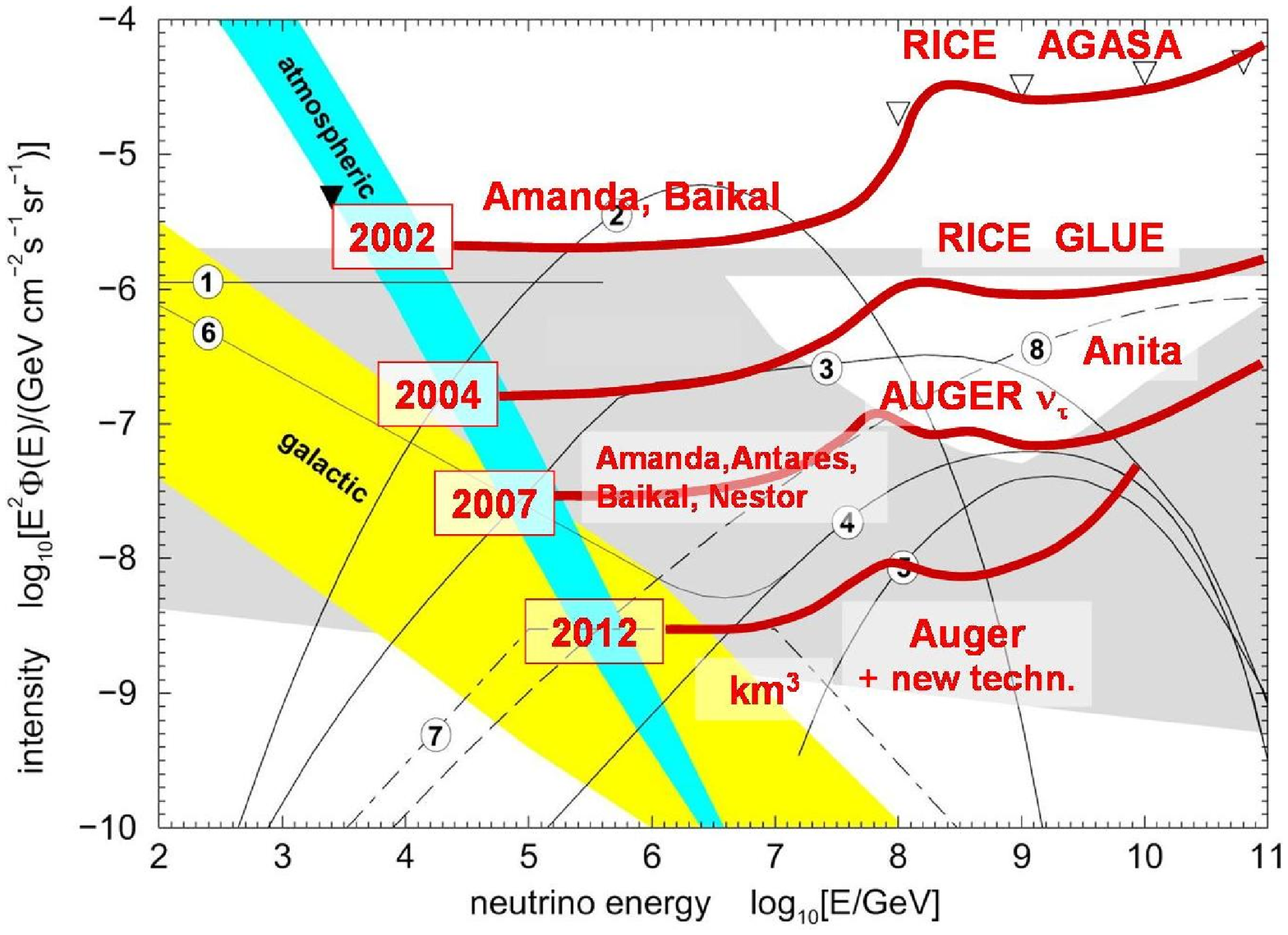,height=10.5cm}
  \caption{Experimental sensitivity to the diffuse neutrino flux for various
           current and future experiments (red lines), compared to different
           models for contributions to the diffuse flux (numbered lines). See
           \pcite{arevns:g29:843} for detailed explanations. The flux of
           atmospheric neutrinos is indicated as blue band. Plot provided by
           courtesy of C.~Spiering.}
  \end{minipage}\hfil
  \label{fig-difl}
\end{figure}

\item
\uline{Search for dark matter annihilation:}\\
The major part of the matter content of the universe is nowadays thought to be
formed by {\it dark matter}, i.e.\ non-baryonic, weakly interacting massive
particles (WIMPs); an attractive WIMP candidate is the lightest supersymmetric
particle, the neutralino. Complementary to direct searches for WIMPs in
cryogenic underground detectors, indirect WIMP observations could also be
possible by measuring neutrinos produced in WIMP annihilation reactions in
regions where the WIMP density is enhanced. Such accumulations may in particular
occur due to gravitational trapping, e.g.\ in the Sun or the Galactic Centre. 

The WIMP signal would be an enhanced neutrino flux from these directions, with a
characteristic upper cut-off in the energy spectrum below the WIMP mass,
$M_\text{WIMP}$. Although there is no generic upper constraint on the
$M_\text{WIMP}$, supersymmetric theories prefer values below $1\tev$. It is
therefore essential for indirect WIMP searches through neutrinos to extend the
detection threshold down to order $100\gev$. The expected sensitivity depends
strongly on assumptions on the WIMP density profile, on $M_\text{WIMP}$ and on
the energy spectrum of neutrinos from WIMP annihilations. At least for some
supersymmetric scenarios this sensitivity is compatible or even better than for
direct searches \cite{astro-ph-0503122}.
\end{enumerate}

\section{Conclusions}
\label{sec-con}

Neutrino astronomy is an emerging field in astroparticle physics offering
exciting prospects for gaining new insights into the high-energy, non-thermal
processes in our universe. The current neutrino telescope projects in the
Mediterranean Sea are approaching installation and promise exciting first data. 
They have reached a level of technical maturity allowing for the preparation of
the next-generation cubic-kilometre detector to complement the IceCube telescope
currently being installed at the South Pole. The interest in this project has
been further enhanced by the recent H.E.S.S.\ observations of high-energy gamma
rays from shell-type supernova remnants in the Galactic plane, indicating that
these objects could well be intense neutrino sources, which would, however, be
invisible to IceCube.

The technical design of the future Mediterranean km$^3$ neutrino telescope will
be worked out in the 3-year EU-funded KM3NeT Design Study starting in February
2006. The construction of the KM3NeT neutrino telescope during the first years
of the next decade thus appears to be possible.

\vspace*{3.mm}

\noindent
{\bf Acknowledgement:} The author wishes to thank the organisers of the 27th
International School on Nuclear Physics, {\it Neutrinos in Cosmology, in Astro,
Particle and Nuclear Physics}, for a very stimulating conference in a unique
atmosphere.

{
\bibliographystyle{./Erice}
{\raggedright\fontsize{10.pt}{11.pt}\selectfont
\bibliography{./Erice.bib}}
}

\end{document}